# Networks: An Economic Perspective

by

Matthew O. Jackson[*], Brian W. Rogers[**] and Yves Zenou[***]

August 2016

## Abstract

We discuss social network analysis from the perspective of economics. We organize the presentation around the theme of externalities: the effects that one's behavior has on others' welfare. Externalities underlie the interdependencies that make networks interesting to social scientists. We discuss network formation, as well as interactions between peoples' behaviors within a given network, and the implications in a variety of settings. Finally, we highlight some empirical challenges inherent in the statistical analysis of network-based data.

[*] Department of Economics, Stanford University, the Santa Fe Institute, and CIFAR, email: jacksonm@stanford.edu, http://www.stanford.edu/~jacksonm.

[**] Department of Economics, Washington University in St. Louis, email: brogers@wustl.edu, http://pages.wustl.edu/brogers.

[***] Department of Economics, Monash University and IFN, email: yves.zenou@monash.edu, https://sites.google.com/site/yvesbzenou.

# 1 Why Should We Study Network Structure?

Why should social scientists care about the patterns of interactions in a society? An easy answer is that science is about explaining what we see in the world and since social structures exhibit regularities we should catalog those regularities and understand their origins. A deeper answer is that, as humans, we should care about the welfare of our societies and social structure is a primary driver of behaviors and ultimately of peoples' well-being. This perspective suggests a series of more directed questions organized around an important agenda. We not only want to understand which regularities social structures exhibit, but also the mechanisms by which social structures impact human opportunities, beliefs, behaviors and the resulting welfare. This also means that it becomes critical to understand why social structures take the forms that they do, and when it is that they take forms that are "better" versus "worse" in terms of the resulting behaviors.

These perspectives lead to questions such as: What is the role of social networks in driving persistent differences between races and genders in education and labor market outcomes? What is the role of homophily in such differences? Why do we see such homophily even if it ends up with negative consequences in terms of labor markets?

## 1.1 Externalities: a unifying theme

The fundamental reason that we need to look at social structure in order to understand human opinions and behaviors is that there are *externalities* involved. "Externalities" is a term that economists use to refer to situations in which the behaviors of some people affect the welfare of others. The concept of an externality is very general and, correspondingly, they can take many forms. Second-hand smoke is a negative externality imposed by a smoker on those around her, as are the $CO_2$ emissions of factories around the world, and some of the numerous activities with positive externalities include acquiring information as well as innovating and inventing. An important aspect of an externality is that, often, the welfare effects on others are not fully taken into account by the decision maker, leading to choices that are inefficient for society. One reason that understanding externalities is important is that they have consequences for policy. For example, as our medical understanding of the effect of smoking has improved, we have learned that the externalities from smoking are significant, which motivated changes to laws restricting smoking rights. In the same vein, the recognition of pollution-related externalities is the prime motivator for regulations such as emissions taxes, which are mostly intended to offset the externalities imposed on others.

Networks of social interactions are important to study and understand precisely because they involve many externalities, both positive and negative. If one person becomes an expert at using a particular software, that knowledge can benefit her colleagues who are then more easily able to learn to use that software themselves. If a person gets a job in a growing company, that position may benefit her friends who may then learn quickly about new opportunities at that company and have an advantage in obtaining interviews. If one business has a partnership with another business that ends up making excessively risky investments in its other dealings and goes bankrupt, then that will damage the partnership and, for instance, if these are large investment banks, could even be a catalyst for a contagion of financial distress.

To summarize, networks are of paramount interest because they drive behavior in contexts that involve systematic externalities. There are two basic themes that arise from this observation.

The first theme involves network formation: people do not necessarily consider the full societal impact of their decisions to form or maintain relationships. This results in suboptimal network formation. For example, a person may not keep in touch with an old friend, even though the information that they might have learned from that old friend could end up being of substantial value to the person's current friends, as may be the case in word of mouth learning about job vacancies. There is a basic sense in which information networks are systematically under-connected from a societal welfare perspective (e.g., see the discussion of the "connections model" below).

The second theme concerns behaviors that involve peer interactions across a given network, such as choices to become educated, to undertake criminal activities, or to adopt a new technology. In these cases external effects lead behaviors involving positive externalities to be taken at too low a level and behaviors involving negative externalities to be taken at too high a level, relative to what would maximize societal welfare.

## 1.2   Overview

The scientific study of social networks was initiated more than a century ago and has grown into a central field of sociology over the past fifty years (see, e.g., Wasserman and Faust, 1994 and Freeman 2004). Although the importance of embeddedness of economic activity in social structures is fundamental, with a few exceptions, economists largely ignored it until the 1990s. Indeed, studies of networks from economic perspectives and using game-theoretic modeling techniques have emerged mainly over the last twenty years (for overviews, see Goyal 2007, Jackson, 2008, 2014; Benhabib, Bisin, Jackson, 2011; Jackson and Zenou, 2015; Bramoullé, Galeotti, Rogers 2016; Jackson, Rogers, Zenou, 2016). By now the study of social networks has also become important in anthropology, political science, computer science, and statistical physics. Network analysis has become one of the few truly multi- and inter-disciplinary sciences. Along with such a variety of fields converging on a common topic come challenges in bridging gaps in terminology, technique, and perspective. Here, we aim to describe some of the perspective and techniques that have come from the economics literature and how they interplay with those from other fields.

Networks are mathematically complicated objects that often involve large groups of people, and thus have daunting combinatorics in terms of their possible connections. Moreover, behaviors and relationships are typically highly interdependent and nonlinear making causal inference problematic. Thus, a scientific understanding of networks and their effects requires the use of models and testing of hypotheses. Which frameworks should we use to generate those hypotheses? One important technique that has emerged from the economic perspective has been the use of game theory.

There are two distinct veins of game theoretic approaches to understanding social networks. The first is analyzing the formation of the network itself. What is the structure of networks that form when the people choose relationships based on costs and benefits? How do the externalities play a role in the extent to which social welfare is optimized? The second vein incorporates an analysis of network structure to help us understand the behavior of agents in a society. This approach includes contagions, diffusion, social learning, and various forms of peer influence. All of these applications involve externalities, and a variety of studies have modeled these various behaviors.

# 2   Network Formation

A presumption underlying the economic modeling of network formation is that people have some choice regarding with whom they interact.[1] Individuals form relationships and contribute effort to maintain them to the extent that they find them beneficial, and avoid or drop relationships that are not beneficial. This presumption is sometimes captured through equilibrium notions of network formation, but is also modeled through various dynamics, as well as agent-based models where certain heuristics are specified that govern behavior. This "choice" perspective traces the structure and the properties of networks back to the costs and benefits that they bestow upon their participants.

It is important to emphasize that this does not always presume that the individuals are fully rational, informed or cognizant of their potential options, or even that they are "individuals" rather than groups or organizations or nations. Nonetheless, the analyses generally embody the idea that networks in which there are substantial gains from forming new relationships or terminating old ones are more ephemeral than networks where no such gains exist.

Regardless of what drives people to form networks, as social scientists we should still care about the welfare implications of the network that forms for a society in question. This interest can be understood from two directions. On the one hand, an analysis of the costs and benefits that networks offer naturally leads us to the question of how well the system performs. Since costs and benefits need to be carefully spelled out, there is a natural metric with which to evaluate the impact of a social network or changes in a social network. On the other hand, some of the interest in social surroundings comes from the fact that models that ignore those surroundings are unable to explain some observed phenomena, such as persistence of inequality among different groups in terms of their employment or wages. The interest in these issues often emerges from some fairness criteria or from an overall welfare perspective, where there is a question of whether or not a society is operating as efficiently as it should or could. For example, when labor markets are viewed in a social context in which opportunities depend in part on one's position in a network, it changes predictions regarding how jobs will be filled and incomes will be determined. This then leads back to a welfarist evaluation of the impact of the network and things like homophily.

---

[1]   With respect to "rational choice" the emphasis is on the "choice" and not on "rationality". Many of the models, in fact, assume that people are quite myopic. The common feature is that people have some discretion in their relationships and make some sort of choices, whether they be biased, myopic, subject to systematic errors, governed by norms, or rational.

Network formation models that account for peoples' choices predict different structures from benchmark models in which relationships are governed purely by some stochastic process that produces a random graph (see Jackson 2008, Newman 2010, and Pin and Rogers 2016, for overviews). There are models that involve choice and chance (e.g., Currarini, Jackson, Pin 2009, 2010; Mele 2013; König, Tessone, Zenou 2014; Chandrasekhar and Jackson, 2014), but the important underlying aspect to modeling choices is that it enables us to account for the externalities present and say something about societal welfare.

Strategic network formation models build on the premise that the payoff or net benefit of a relationship depends potentially on the larger network, and are not just on the dyad in question. For instance, people obtain value from friends of friends, and so have an interest in the way in which the network is connected that extends beyond their immediate neighborhood.

Game theory provides a set of tools to analyze such a situation. Strategic network formation analysis borrows ideas from this literature and has also contributed to its recent enlargement. Nash equilibrium is a central concept in game theory[2] but it turns out to be a clumsy tool for modeling network formation as there often needs to be *mutual consent* in link creation. A person cannot simply decide to be a friend or partner of another; it requires both to agree. To get around the fact that Nash equilibrium embodies only unilateral strategic adjustments, one could try to model a process of proposals and acceptances, but it becomes needlessly complicated and hard to analyze.

---

[2] Indeed, in game theory, *Nash equilibrium* (named after John F. Nash) is the canonical solution concept of a game. It is a list of a strategy for each player such that no player has anything to gain by changing his or her own strategy (i.e., unilaterally). For a brief introduction to the basic definitions and workings of game theory, see Jackson (2011).

To handle such mutual consent in network formation, Jackson and Wolinsky (1996) introduced an alternative solution concept for network formation games, which they referred to as *pairwise stability*.[3] To be pairwise stable, a network has to satisfy two conditions. First, no agent can benefit from severing any of his or her existing links. Second, no pair of agents should both benefit from creating a new link between them. Pairwise stability is a relatively weak equilibrium or stability concept that embodies mutual consent.

Jackson and Wolinsky (1996) also introduced an illustrative model that they called the *connections model*, which captures values of indirect connections and externalities in a network. People benefit from friendships, but they also benefit from friends of friends, and from friends of friends of friends, and longer paths in the network. For instance, having more friends of friends may give a person more indirect access to information as it flows to the person's friends. In particular, any connection, direct or indirect, is valuable, although more distant connections contribute less to an agent's well-being. Direct relationships also involve maintenance costs. People weigh the costs and benefits of their friendships, including the indirect benefits that they bring. They form valuable relationships and not ones that would be of a net negative value. A variety of network structures can be pairwise-stable and, quite naturally, that set depends on the costs and benefits of forming relationships, as well as the values of indirect relationships. To see some of the basic ideas underlying the literature, consider the following star network (Figure 1), with a center (agent 1) connected to three peripheral agents (agents 2, 3, and 4).

**Figure 1: A star network with four agents**

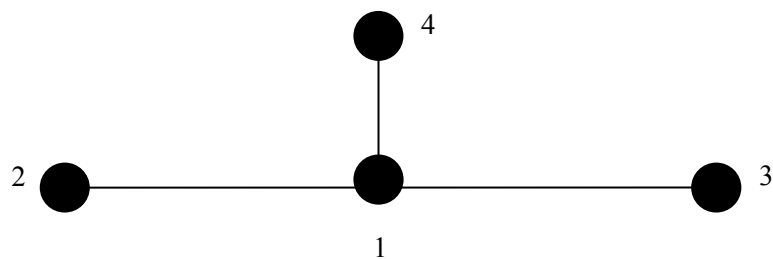

---

In any model in which agents benefit from indirect connections, such as the connections model, the network that forms might not be the one that results in the highest social welfare. For instance, in Figure 1, there may be some indirect benefit that agent 2 gets from having agent 3 as a "friend-of-a-friend". Agent 1 will generally not take this indirect benefit for agent 2 fully into account (even if partly altruistic) when deciding whether to maintain a relationship with 3. This positive externality means that private incentives and overall total social welfare are misaligned. In the case in which there are positive externalities, unless all agents are exactly and fully altruistic, the networks that form will generally not be the ones that are best from a societal point of view.[4]

There are also many settings in which ties result in negative externalities. For instance, it could be that agent 1 is collaborating on a research project with 3 and that distracts agent 1 from spending time with agent 2. In that case, there is a negative externality on agent 2 that agent 1 may not fully internalize when forming the relationship with agent 3.

Such analyses reveals that, in general, individual and social incentives are not congruent, and that decentralized behavior typically results in an inefficient arrangement of social ties. A basic intuition is that with positive externalities too few social ties are formed, as people do not fully internalize the indirect benefits to others when forming links; and conversely, with negative externalities too many social ties are formed as people do not fully internalize the indirect harm that their relationships do to others. Indeed, for a wide variety of settings this is the case (e.g., see the discussion in Jackson, 2003).

It is not easy to overcome the tension between individual incentives and societal welfare due to the pervasive externalities in network settings. To some extent, people who provide indirect benefits to others may expect to be compensated for their behaviors. For example, Burt's (1992, 2004) theory of structural holes relates to this idea: people who connect otherwise disparate groups benefit from those connections. This is not solely because they benefit directly personally from their key position, but also because they serve as an important conduit and coordinator – providing externalities to those for whom they serve as a connector. They may be able to leverage that key position, for instance, by expecting favors in return for the unique information that they are able to relay to others. Although such rewards may help induce people to provide critical connections and positive externalities, the rewards generally cannot reflect the full value of all of the externalities present in a network at the same time, and thus fall short of providing the full incentives needed to result in networks that are best from society's perspective. This can be shown mathematically.[5]

---

[4] This applies with various measures of overall social welfare, including a utilitarian measure, as well as Pareto efficiency. See Jackson (2003) for a discussion.

This tension between individual incentives and overall societal efficiency has been examined in wide variety of models of costs and benefits, ranging from stylized versions of costs and benefits from friends of friends to more explicit models of things like Research and Development Networks. One avenue pursues the case in which ties are directed and can be formed unilaterally (e.g., see Bala and Goyal, 2000), for instance as in citations or in linking to or following on various social media. For such settings, externalities are again present and similar results apply as in the case with mutual consent in tie formation. A second avenue has considered heterogeneous versions of the connections model, wherein the costs and benefits of forming links can vary across agents (and pairs of agents), see, e.g., Johnson and Gilles (2000), Jackson and Rogers (2005), Galeotti, Goyal and Kamphorst (2006), Carayol and Roux (2009), and De Marti and Zenou (2016).

In all of these models characterizing the full set of equilibria can be challenging given the combinatorics. Cabrales, Calvó-Armengol, and Zenou (2011) (see also Currarini, Jackson, Pin, 2009, 2010) propose an alternative approach where agents only decide upon the socialization effort they exert. For example, researchers go to workshops and congresses to listen, to be listened to, and to meet other researchers. In this approach, socializing is not equivalent to elaborating a nominal list of intended relationships and, more importantly, meeting and talking to someone does not necessary implies forming a link with that person. Indeed, if we take the example of conferences, then individuals decide how often they go to conferences. The probability of forming a link between two persons will then positively depend on how often these two persons go to the same conference. But, even if these two persons go very often to the same conferences and talk a lot to each other, it does necessary imply that they will write a paper together (i.e., form a tie). Cabrales Calvó-Armengol, and Zenou (2011) can then perform an equilibrium analysis that equates marginal costs and benefits of socialization, and characterize the equilibrium and examine its welfare properties. They show, again, that externalities lead the equilibrium socialization efforts to be inefficient.

---

[5] See Jackson and Wolinsky (1996), Dutta and Mutuswami (1997), Currarini and Morelli (2000), Watts (2001), and Bloch and Jackson (2007) for more discussion, and see Jackson (2003, 2008) for overviews.

More recently, researchers have been developing richer dynamic and statistical models of network formation (e.g., Jackson and Watts (2002), Jackson and Rogers, 2007; Christakis, Fowler, Imbens and Kalyanaraman, 2010; Mele, 2013; Chandrasekhar and Jackson, 2014, 2016; König Tessone, Zenou, 2014; Badev, 2015; Leung 2015; De Paula, Richards-Shubik and Tamer, 2015; Graham, 2016) that can incorporate incentives to form relationships and are suited to empirical application and statistical analysis.

# 3  Behavior and Games on Networks

Thus far our discussion has focused on network formation. In this regard, the analysis typically builds from a given specification of how a network, once formed, translates into the costs and benefits for each agent. For the most part, those models do not derive the payoffs to the people involved in the network from a comprehensive look at how people behave once a network is in place. For example, the connections model posits that friends-of-friends are valuable, but does not model that value as coming from a specific social learning or diffusion process. One can imagine different kinds of interactions that may ultimately correspond roughly to the payoffs specified in the connections model, but the interactions themselves lie outside the description of that model.

This leads to a question of how people will behave once a network is formed, and how that depends on the setting and the externalities that are generated by their behaviors. In the present section, we discuss the literature that provides a complementary view to strategic network analysis, by explaining how behavior is driven by incentives and externalities on a given network (and so does not attempt to model network formation). This is then more explicit about the details of how agents interact along their links and what drives those behaviors. Again, a main question of interest is how externalities distort behaviors away from what might be socially optimal.

There are many simple interactions, including basic forms of contagion and diffusion processes that are studied in this way without an explicit consideration of incentives. In fact, there is a very large literature discussing such processes on networks. For example, there is a wide class of models developed for epidemiologic questions concerning how a disease may spread through a population (see e.g. Bailey, 1975 and Diekmann, Heesterbeek, Britton, 2013). Important questions there include understanding when a disease will take hold in a society once introduced versus when it will eventually die out. In the latter case, how much (and which members) of a society will be infected, and so forth. It has been shown that answers to these questions depend in important and systematic ways on the structure of the contact network as well as how the spreading depends on the level of exposure (see, e.g., Granovetter 1978; Pastor-Satorras and Vespignani 2001; Lopez-Pintado 2008; Jackson and Yariv, 2007, 2011; Jackson and Lopez-Pintado, 2013). The role of externalities in such settings also readily apparent. For example, a choice by one agent to become immunized against a disease confers a positive externality on those which whom she interacts.

For many behaviors, however, choices of people interact with the decisions of their friends. For instance, a decision of whether to adopt a new technology is usually driven by whether those around the decision maker are using compatible technologies. This means that such a diffusion process is more complex than the propagation of a disease. Correspondingly, these models tend to be naturally considered through game theoretic approaches, wherein each agent is presumed to take an action from a specified set, and receives a payoff that depends both on her own action and the actions chosen by other agents − e.g., her friends and acquaintances in the network. Here externalities are pervasive and this applies to most peer-effects settings.

Note that, even in the usual case in which payoffs depend on the actions only of direct neighbors in the network, agents will care about, and have to make inferences regarding, the actions of further away agents. Friends of friends may not matter directly, but they do influence the behavior of one's friends. This means that indirect effects can be of prominent importance in equilibrium characterizations.

We refer to this literature as that of "games on networks" (for an overview see Jackson and Zenou, 2015). We emphasize that a very wide set of applications fall under this umbrella, including the diffusion of any good that involves complementarities, local public goods production, and any peer-effect application. While the payoffs can generally take an unrestricted form, two important classes of network games have been identified that give rise to important general insights. The classes are, respectively, games with strategic complementarities and with strategic substitutes.

To understand the essence of these classes of games, consider a game where each agent chooses some level of behavior: for instance how much eductation to obtain, or how many hours to study, or how hard to work, or how much effort to put into learning how to do something, etc. Strategic complementarity describes the case in which an increase in the behavior of a friend *increases* the marginal benefit from increasing one's own action. For example, the marginal returns to working on a collaborative project may increase in the time that one's partners put into the project due to synergies. The equilibria of games with strategic complements are generally those for which, as one agent increases her action, her neighbors find it in their interest to increase their actions, which then feeds back to create an additional incentive for higher actions to the original agents, and so on. Note that even with such positive feedback, this does not necessarily lead to socially optimal outcomes, since each person is still choosing a level of action based at least partly on their own well-being and not based on the overall impact on the society.

Games with strategic substitutes are, in contrast, those in which an increase in the action of a friend *decreases* the marginal incentive to raise one's own action. For example, as one's contacts vaccinate themselves against a disease, the incentive to immunize oneself decreases, as the chance of contracting the disease is lowered by the actions of one's neighbors. Roughly speaking, these settings games tend to have a less "extreme" nature, since higher actions by some agents tend to produce lower actions in connected agents. There are still feedback effects, but here instead of amplifying behaviors, the feedback is to mitigate the behaviors. Again, there are still important externalities, since any one person's decision is still impacting many others, and the individual is not always weighing the full social impact of his or her own decision.

## 3.1　Strategic complementarities

To illustrate the characteristics of network games with strategic complements let us examine a specific and very tractable game in which links represent the exertion of influence between two peers in a social network.

To that end, consider a set of agents who each choose how intensely to pursue a social activity. The returns to investing in the activity depend on the agent's action and the actions of her neighbors. A fixed network governs who influences each given agent (e.g., her friends). The game is one complements, in which an agent's marginal payoff from the social activity is an increasing function of the choices of her neighbors. We focus on an interpretation of the activity as criminal behavior, so that as one's friends increase their level of criminal activity, the returns to increasing one's own criminal activity increase as well (although other interpretations are possible as well, such as investment in education).

Ballester, Calvo-Armengol, and Zenou (2006) analyze such a game where the utility functions take a particular "linear-quadratic" form. They characterize the Nash equilibrium of this peer-effect game. A central finding is that an agent's criminal activity is proportional to how central she is in the underlying network, where centrality is measured according to a particular formulation known as "Katz-Bonacich" centrality (a well-known measure introduced by Katz, 1953, and Bonacich, 1987). The relationship between the peer effect and Katz-Bonacich centrality is quite intuitive, as they both involve iterative interactions. A person is influenced by her friends. They are in turn influenced by their friends. Thus, friends of friends have an indirect effect and that effect can also be quantified. Iterating, there are effects of friends of friends of friends, and so on. Katz-Bonacich centrality looks at this limit, where there is a decaying effect as the distance increases, and this is exactly how such games with complementarities work as well, and so it is very natural that this centrality measure plays a prominent role in understanding peer-effects.

What kinds of insights can be taken from this model? As just mentioned, the analysis predicts that a key determinant of one's (e.g., criminal) activities is the position in the network: the more central (in terms of Katz-Bonacich centrality) a person in a network, the higher is his/her level of criminal activity. The centrality captures the direct and indirect influences on a given individual. This gives a causal prediction between network position and criminal activity, and here Katz-Bonacich centrality embodies the externalities and how they operate through the network.

Analyses of such models also yield further predictions. For example, one can also use them for policy experiments, seeing how changing the network or intervening to influence behaviors can change the overall behaviors in the society. In particular, in the context of the model mentioned above, it suggests a way of formulating policies aimed at reducing crime. Indeed, Ballester, Calvo-Armengol, and Zenou (2006, 2010) propose a new measure of network centrality: *inter-centrality*. This measure can be used to identify *key players* in a network; i.e., those people whose removal from the network would lead to the greatest aggregate reduction in criminal activity. Inter-centrality ranks individuals according to this criterion, and allows one to identify the most effective targets when one's objective is to reduce overall crime.

Importantly, Katz-Bonacich centrality and inter-centrality need not rank individuals in the same way. One captures how a given person is influenced by others, while the other captures the overall influence of a person on the society's activity. It can be that the most active criminal is not the most effective target for reduction of overall crime. This potential difference derives from the complicated set of externalities implicit in peer-effect settings, and the fact that how a person is influenced is not the same as how influential that person is on others.

To illustrate this finding, consider the following network of 11 criminals:

**Figure 2: A bridge network with 11 agents**

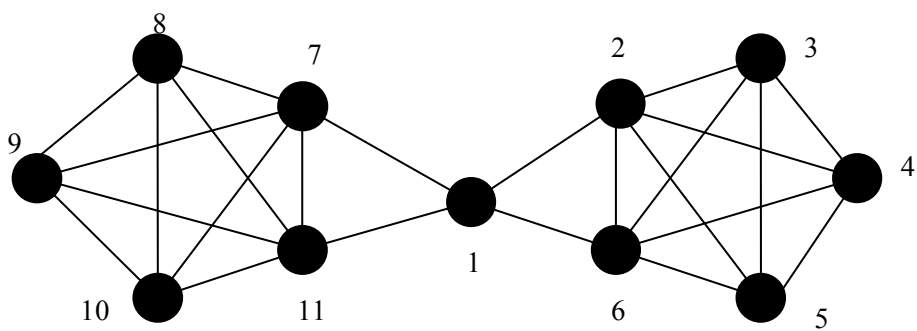

Let us determine who the most active criminals and the key players are. Actually, these depend on a factor $0 < \delta < 1$ that discounts how influence decays with distance (e.g., how much influence does my friends' actions have on my activity). Similar to the connections model discussed earlier, the influence of people at a distance $k$ from an individual on that individual are proportional to $\delta^k$ so that criminal 4's influence on criminal 11 is proportional to $\delta^3$. Calculating Katz-Bonacich centrality and inter-centrality, both make use of this discount factor. It turns out that, whatever the value of $\delta$, individuals 2, 6, 7 and 11 display the highest Katz-Bonacich centrality. These are the individuals who have the highest number of direct connections and, importantly, they are directly connected to the "bridge individual" 1, which gives them relatively close access to a large range of indirect connections. Altogether, they are the most central criminals and thus, in equilibrium, those who engage most heavily in crime.

For small values of $\delta$ (e.g., $\delta$=0.1), the key player is also the most active player (i.e., of type 2) but for higher values of $\delta$ (e.g., $\delta \geq 0.2$), the key player is criminal 1, the bridge player. Even though that person does not have the highest Katz-Bonacich centrality, removing that person from the network would reduce crime by the most since they are most essential in transmitting behavior across the network. Indeed, because indirect effects have a pronounced effect, eliminating criminal 1 has the highest joint direct and indirect effect on aggregate crime reduction. This is only true when there are enough externalities from removing criminal 1. When $\delta$ is small enough, indirect effects are small and direct links drive behavior. As a result, criminal 2 who has more direct links than criminal 1 becomes the key player.[6]

Generally, in networked peer-effects and games and network settings, whether the behaviors are larger or smaller than socially optimal depend on two aspects. The first is whether behaviors are complements (e.g., studying, criminal behavior, cheating) or substitutes (e.g., local public goods, vaccinations), as that dictates whether behaviors are amplified by the behavior of others or muted by the behavior of others. The second is whether the externalities are positive or negative. That is, do the behaviors of others impact a given person positively or negatively. It is important to emphasize that this is very different from whether behaviors are complements or substitutes. The fact that behavior is strategic complements in both the case of education and crime, leads to similar effects in terms of their feedback, but very different policy implications. In the case of studying, there is suboptimal behavior. By subsidizing behavior, or making it easier for students to study, a government could increase social welfare. In contrast, in the case of criminal behavior it is exactly the opposite: even though the game is still one of complements, the overall external effects are negative – criminal behavior ultimately harms others.

---

[6] Using data on American teenagers and Swedish adult criminals, Liu, Patacchini, Zenou and Lee (2012) and Lindquist and Zenou (2014) show, respectively, that, in the real world, 20-30% of the most active criminals are not the key players. For an overview on the literature on key players, see Zenou (2016).

The literature in this area has been successful in outlining the impact of externalities in both games of complements and substitutes, and how these apply to a variety of settings (again, see the survey by Jackson and Zenou, 2015). Some related areas in which people make decisions and these interact are discussed below. These are just a few of the application areas, and provide some illustrations of the insights that emerge.

## 3.2    Financial Networks

Financial markets are a setting in which networks of relationships play a central role, and in which there are relevant and timely problems for which network analysis is essential. Systemic risk in financial settings is a network phenomenon. There are many dimensions on which network externalities drive financial fragilities, and here we mention two of them. One relates to panics and bank runs, and the other relates to counterparty risk, which has to do with the chance that a business partner will default which is driven by their other investments.

The role of networks in financial panics is well-illustrated by a fascinating historical study of Kelly and Ó Gráda (2000) who look at the behavior of Irish depositors in a New York bank during two bank-runs in the 1850s. As recent immigrants, their social network was determined largely by their place of origin in Ireland, which largely determined where they settled in New York, and thus their social networks. During both panics this social network was the prime determinant of if and when a depositor pulled their money from the bank. In those times, even some financial market information was largely by word-of-mouth, and so the panic spread according to classic diffusion patterns, with basic insights applying from the classic contagion literature.

Although classic network analysis applied to such word-of-mouth panic spreading, there are additional subtleties that drive systemic risk in financial networks. There is a rapidly growing literature on this subject that merges classic network analysis with details from financial settings.

The starting point is the obvious one that the health of a financial institution derives, in part, from the health of other financial institutions. Links, representing contracts and liabilities between institutions, allow for the possibility of contagion, in that if one institution becomes insolvent, this can cause otherwise healthy partner institutions to become unhealthy. This can cascade and have a large effect on the economy. Even a relatively small shock to the system can become amplified and have profound effects on the performance of a whole economy.

To understand this issue, consider banks in different regions (e.g., as in Allen and Gale, 2000) that hold part of their deposits in other regions for some diversification. In the case of a banking crisis in one region, the claims of other banks on this region lose value and can cause a banking crisis in the adjacent regions. The crisis can gain momentum as losses in those adjacent regions spread in a contagion to more regions.

Here, however, there is an important tradeoff that makes such contagions different from the standard ones that we are used to from epidemiology and from marketing studies of diffusion. In financial networks there are countervailing effects. As financial institutions and/or regions become more inter-connected or globalized, the network possibilities of contagion multiply. One the other hand, as the institutions or regions become more inter-connected they also become more diversified and less sensitive to shocks to any particular partner. Which of these effects dominates depends on the specific context. As long as the magnitude of shocks affecting financial institutions are sufficiently small, denser networks enhance the stability of the system (Allen and Gale, 2000; Freixas, Parigi and Rochet, 2000). However, as shocks become larger, such interconnections serve as a mechanism for propagation of shocks and lead to a more fragile financial system. Which sort of shocks matter depends on the both the topology of the network and the size of the contracts that are present in each link (Elliott, Golub and Jackson 2014; Acemoglu, Ozdaglar, and Tahbaz-Salehi 2015).

This is important because it helps us understand the most recent financial crisis in which there was a relatively large shock to the system (a crisis in the mortgage market). The well-connected network of swaps and other contracts led financial distress to spread widely. This led too-big-to-fail considerations to evolve into too-interconnected-to-fail concerns and increased attention to the understanding of how networks matter in financial contagions. The full extent of the network of relationships was under-estimated, as was the correlation in the investments and the extent to which the mortgage crisis could reach.

Here, again, externalities play a central role. One bank that has a financial contract with another cannot control what risks the second bank takes with the rest of its portfolio, and may not even be able to assess that counterparty risk. The second bank is making investments based on its own tolerance for risk and may not pay attention to the consequences it will have for the larger economy if it is forced to default. The managers of Lehman Brothers were not necessarily paying attention to the future costs to shareholders of other institutions and to taxpayers that their bankruptcy would trigger, as hundreds of billions of dollars needed to be pumped into the system to prevent a major contagion. Understanding such financial networks and systemic risks is an important area of active research.

## 3.3   Social Learning

Much of what we understand about the world is learned from observing others. These inferences depend on what others know about the world and also potentially on what decisions they make. We rely heavily on others' reviews of products and we make inferences about quality from the popularity of goods and services. These outcomes then involve externalities, as others' decisions influence our inferences and decisions, and ultimately our welfare. While this simple observation has always been relevant, it is becoming increasingly so as more information about others' likes and dislikes are communicated via online systems.  There is a vast literature on social learning spanning a number of disciplines, but a couple of key papers are very useful in pointing out the externalities in such settings and their potential impact: Banerjee (1992) and Bikhchandani, Hirshleifer, and Welch (1992). These seminal papers study a simple framework within which individuals have privately held information about the quality of some goods (for example, the quality of a restaurant). While this information cannot be directly communicated to others, individuals sequentially take publicly observed actions based, in part, on their private information (e.g., which restaurant to patronize). If properly aggregated, the private information would reveal the quality of the good almost perfectly in a large population, but the externalities are such that this social learning may fail.

A central message of these papers was to emphasize that, under certain conditions, *herds* are inevitable: after some point, all individuals take the same action, even if it is an inefficient action. The basic mechanism is fairly simple, so we illustrate the intuition and refer the reader to the papers for more details. Consider people deciding whether to buy a new product or not. Consider a very simple case in which everyone would actually agree that the product is not worth buying if they had a chance to try it out. People look at the product and get some impression of it, and based on these impressions have some information about its quality. The key assumption is that any one person could be wrong, but on average they would be right: if we pooled the population, the majority of people would correctly infer that the product is not a good one. However, people move one by one. Consider a case in which the first person has a good impression and buys the product. Suppose, by chance, that the second person also happens to have a good impression and buys the product. Now consider the third person. Even if she has a bad impression, she sees that the first two people bought the product and so infers that they both had good impressions. She then realizes that the majority of observable impressions are positive and so she buys the product, *regardless of her own impression*. The fourth person therefore learns nothing about the third person's impression and thus faces the same decision problem as the third person and also ignores her own impression. A herd forms and everyone buys the product even though it is a dud.

Clearly this herding effect is the result of a very stylized model, and whether it holds or not depends on details of the setting (e.g., see Smith and Sorensen, 2000), including the quality of information, heterogeneity of the population, whether people are altruistic (and for instance, who posts reviews on line), and who sees whom.

These models are set up to highlight the role of *informational* externalities. The payoff to each individual depends only on her own decision and the quality of the product; and not on the action taken by anyone else. However, peoples' actions are still important for others as they carry information. People do not account for the informational content of their actions for others when making their decisions. If the third and fourth people in our example's queue had followed their own impressions it would have given more information to the subsequent society. Instead, the selfish act of following the herd, while individually rational, confers an informational externality to subsequent decision makers, since they suffer from not being able to learn from many of the actions before them.

Again, although this model is stylized, it is clear that people are influenced by others when making decisions. For example, in the classic study on the Truman versus Dewey presidential campaign, Lazarsfeld, Berelson and Gaudet (1944) found that voters were more influenced by friends and colleagues rather than by the mass media. Given the widespread applications, and the fact that the externalities become even more complex in richer settings, the literature on social learning has expanded in many directions. Social learning can fail in many ways and this had led to many models of social learning in networks, in which the structure of who observes the actions of others, and how their actions are sequenced, is more elaborate. There are, in fact, two main paradigms in the literature, often referred to as *rational (Bayesian) learning* and *naïve (boundedly rational) learning*, respectively.

The main idea behind the Bayesian learning model is that agents are fully rational, in the sense that they accurately process any information that becomes available to them, either through communication or through the observation of the actions and the payoffs of their peers. As (Bayesian) consistency would suggest, under some regularity conditions, agents in large networks will converge to the same beliefs and/or actions, *but not always to the optimal ones given the externalities* (e.g., see Bala and Goyal 1998; Tahbaz-Salehi and Jadbabaie, 2008; Mossel, Neeman and Tamuz, 2014; Mossel, Sly, and Tamuz, 2015; Molavi, Eksin, Ribeiro and Jadbabaie, 2016; Golub and Sadler, 2016).

In the naïve (boundedly rational) learning model, introduced by DeGroot (1974), agents start with some initial beliefs and then update their beliefs by taking a weighted average of their neighbors' beliefs. Agents are boundedly rational in the sense that an agent's new belief is just the (weighted) average of his or her neighbors' beliefs from the previous period which, e.g., can lead to "over counting" information from some individuals, relative to Bayesian updating. Over time, provided that the network is strongly connected (so there is a directed path from any agent to any other) and satisfies a weak aperiodicity condition, beliefs converge to a consensus. It can be shown (Golub and Jackson, 2010) that, only if the influence of prominent individuals goes to zero as the network grows, society can learn efficiently.

The Bayesian approach hinges on the assumption that agents possess the mental capacity to optimally extract and aggregate information in the aforementioned way, or at least along similar lines. Although in some cases this assumption may be plausible, empirical evidence suggests that even in very small and simple network structures people are far from rational in their processing of others' information (Choi, Gale, Kariv, 2008). In fact, observations from recent field experiments (Grimm and Mengel, 2015; Mobius, Phan and Szeidl, 2014; Chandrasekhar, Larreguy and Xandri, 2015) are compatible with the assumption that subjects exhibit a non-Bayesian behavior more compatible with the naïve (boundedly rational) learning model.

There are several ways in which externalities manifest themselves here; let us mention a few of the ways in which externalities lead learning to fail. First, as we have already seen, it may be that agents are not taking into account the information value of their actions to others. Also, it can be that agents don't experiment as much as would be optimal. For instance, consider our example of buying the new product above, but suppose that people also see previous peoples' satisfaction, and not only their decisions: for instance, seeing reviews of previous purchasers. Here it can be that too few people try the product for society to end up learning the product's quality. Even if it is not in some particular person's interest to try the product because the current information suggests that the product is not worthwhile, it could be best for the overall society to have that person experiment and try the product in order to get better information and be more confident that the product is not worthwhile. This follows since if it turned out to be a good product, it will generate positive value to many other people – and so society would like to see more experimentation than any given potential consumer would.

Another possible failure in social learning comes not from the particular decisions of the agents, but from the structure of the network. For instance, it might be that people pay too much attention to just a few people (Bala and Goyal 2000; Golub and Jackson 2010) or have excessive homophily (Golub and Jackson, 2012). Indeed, homophily (the tendency of individuals to associate with individuals with similar characteristics; see McPherson, Smith-Lovin, and Cook 2001) can reduce the speed social learning: opinions may converge quickly within a group, but may move slowly across groups. Individuals may prefer to form ties with similar others but this could be harmful to the overall ability of society to process information.

## 3.4   Labor Markets

A wide set of studies documents that the filling of most jobs involves, at least in part, personal contacts (e.g., see Rees, 1966; Rees and Shultz, 1970; Granovetter, 1973, 1974; Holzer, 1988; Ioannides and Loury, 2004; Wahba and Zenou, 2005; Bayer, Ross and Topa, 2008; Pellizzari, 2010; Topa, 2011; Kramarz and Nordström Skans, 2014; Brown, Setren and Topa, 2016). Indeed, networks of personal contacts are critical conduits of information about employment opportunities, which flow via word-of-mouth and, in many cases, constitute a valid alternative source of employment information to more formal methods. Networks of contacts have the advantage that they can be relatively less costly to use to identify appropriate workers than sorting through endless numbers of applications; and they can provide more reliable information about jobs and the workers, and hence lead to better matches, than other methods.

Given that networks place constraints on who hears about job opportunities, the classical "frictionless" supply and demand analyses of labor markets has the potential to be quite misleading. Rather, which firms hire which workers, and how those workers are compensated, depends in important ways on how information about job vacancies is passed through society. Thus, there is an immediate sense that being "well-connected" confers advantageous opportunities to learn about attractive jobs, and improves an individual's chances of employment, the quality of their jobs, and their compensation.

Indeed, such an intuition is borne out in models that account for word-of-mouth learning about job opportunities. Better-connected individuals have shorter periods of unemployment and higher wages on average. In fact, modeling such a setting, Calvó-Armengol and Jackson (2004, 2007) explore wider implications social networks in labor markets. For example, they show that if there is some homophily or segregation in a society, and one begins with any disparities in employment across segments of the population, then those disparities can be amplified and become persistent. If one's social contacts are poorly employed, then they are a less reliable source of information for several reasons, including the fact that they are less likely to hear about job opportunities at their employer (since they frequently have none) and also they become competitors for some job opportunities that do become available. Thus, people who have better-employed friends and relatives have greater incentives to stay in the labor market and to invest in education and human capital. Given this prediction, if there are segments of the population that are poorly-employed and others that are well-employed, those disparities can be robust and persistent. In terms of social well-being, to the extent one is concerned with reducing inequality (at least of opportunities if not also the outcomes themselves), an important message is that policies aimed at distorting labor market outcomes must account for the correlations and persistence in labor markets, and associated education decisions, induced by job contact networks.

In this framework, externalities are again important in understanding the conclusions. For instance, when one worker drops out of the labor force, she imposes a negative externality on her friends and relatives who could have benefited having an (employed) friend as a contact. These frictions further lead to mismatches in the labor market, as there is no guarantee that the best worker for a given job will be matched to that job, as the optimal worker may never learn about the opening or acquire the skills needed. Recent empirical work examines a variety of aspects of the distortions in labor markets (see e.g., Beaman, 2016, for an overview).

## 3.5    Development Economics

When formal institutions do not exist or do not work very well, informal, network-based relationships serve as substitutes. For example, when insurance markets are missing and access to banking is limited, shocks to incomes are often smoothed via borrowing from family and friends (see e.g.., Fafchamps and Lund, 2003; Kinnan and Townsend, 2012). Further, new technology adoption is critical for growth and development (see e.g., Alvarez, Buera and Lucas, 2013; Perla and Tonetti, 2014), and social relationships are the means by which many individuals learn about, and are then convinced to adopt, new technologies. These and other vital functions of social networks have made them a main focus of development studies.[7]

For example, the adoption of new technologies and programs have been the focus of many studies such as, for example, Foster and Rosenzweig (1995), Conley and Udry (2010), Duflo and Saez (2003), Banerjee, Chandrasekhar, Duflo, Jackson (2013, 2016), Beaman, BenYishay, Magruder, Mobarak (2016). Banerjee, Chandrasekhar, Duflo and Jackson (2013) use network data to analyze the diffusion of microfinance in a set of rural Indian villages. One of their key questions is: How do the social networks in a village affect the diffusion of microfinance loans? They develop a model of information diffusion through a social network that discriminates between *information passing* (individuals must be aware of the product before they can adopt it, and they are made aware by their friends) and *endorsement* (the decisions of informed individuals to adopt the product might be influenced by their friends' decisions). Their findings suggest that households who participate in microfinance are much more likely to inform their friends that microfinance is available than households who choose not to participate. Thus, households who have a high fraction of friends who are participating are much more likely to hear about microfinance than those with a low fraction, all else held constant. Their analysis suggests that much of the peer interaction in this setting involves people making each other aware of microfinance and that peer influences beyond that play an insignificant role in participation decisions. Moreover, they trace the wide variation in adoption across villages to the centrality of the first-informed households in the villages, and show that how one measures centrality is essential to predicting diffusion.

Beamon, BenYishay, Magruder, Mobarak (2016) study a similar issue via a carefully designed randomized controlled trial of program that incentivizes farmers to adopt a productive new agricultural technology in 200 villages in Malawi. Their aim is to identify seed farmers to train to use the new technology. They find that some farmers need to learn about the technology from multiple people before they adopt themselves. Thus, in that context, simply changing the patterns of who is trained in a village on a technology on the basis of social network theory can increase the adoption of new technologies compared to the Ministry's existing extension strategy.[8]

---

[7] For recent overviews, see Fafchamps (2011) and Breza (2016).

Failures of diffusion in some of these settings relate back to the externalities that we discussed in the context of social learning. Here, the choice of whether to adopt a new technology by one household ends up affecting both whether their neighboring households become aware of the technology and also whether those neighbors become convinced to adopt it. Diffusion can fail for both of these reasons, and these studies together show that it can be vital to distinguish and account for both effects as they differ across products. In some cases a new technology is easy to understand and clearly beneficial and so failures in diffusion come from network structures and failures in the spread of information, while in others additional learning is required and additional externalities are present as people learn from the experiences and behaviors of others.

## 3.6    Exchange Theory, Bargaining, and Trade on Networks

Another obvious area in which network structure impacts outcomes, and in which people exercising power in key network positions impacts the terms of trade that others obtain, is that of the exchange of goods and services. Such trade, along with the associated bargaining processes, rarely occurs in the context of a centralized market; much more often it occurs via a series of bilateral agreements. When individuals who are critical to certain transactions (e.g., see Burt, 1992) exercise their bargaining power to obtain favorable terms, this can lead to substantial frictions in trade and to inefficient transactions and allocations in the system.

Early experiments, such as those by Cook and Emerson (1978) and Bienenstock and Bonacich (1993) show how position in a network affects which bargains are struck and how gains from trade are distributed across a network (see Cook and Cheshire, 2013, for a recent survey). Game theory has played an important role in such analyses including early studies by Aumann and Myerson (1988) and Bonacich and Bienenstock (1993), and has driven the rapidly expanding literature that spans theory and experiments.

---

[8] This is related to the experiments of Centola (2010, 2011) who finds that the clustering in network structures can affect the adoption of some products.

Indeed, in many markets a buyer and a seller must make an investment before they can (feasibly or profitably) trade. This may involve setting up accounts, opening channels of communication, making personal contacts, or making goods and systems compatible. This is true of a variety of markets from the production of clothing (e.g., see Uzzi 1996) to the trading of securities (e.g., Wang 2016). The market is networked due to the constraints imposed by the (lack of) investments. The network structure of possible trading pairs then becomes relevant because it determines the outside options of a given buyer and a seller bargaining over a trade surplus. For example, a buyer who has invested with only one seller may have little bargaining power in that relationship, since she has no alternative partners with which to profitably trade. Kranton and Minehart (2000), Jackson (2003), Corominas-Bosch (2004), Charness, Corominas-Bosch, and Frechette (2005), Blume, Easley, Kleinberg and Tardos (2009), Manea (2011), Elliott (2015), Nava (2015), Condorelli, Galeotti, and Renou (2016), Wang (2016), among others, analyze various market structures and models. For example, in addition to pure buyer-seller networks, trade can more generally involve intermediaries, dealers, brokers and retailers, whose motives to trade are pecuniary rather than deriving from final consumption of the traded goods. The presence of intermediate traders raises a set of questions: how do different network structures affect the efficiency of trading outcomes? How does the position of a trader in a network affect her profit? (For overviews, see Cook and Cheshire, 2013, Condorelli and Galeotti 2016, and Manea 2016).

There is also a large literature on inefficient investment due to hold-up problems[9] and overinvestment in outside options. For example, Elliott (2015) shows that there are inefficiencies in these investments when long-term contracts cannot be enforced. One insight that emerges here is that a given agent can have a strong incentive to make costly investments in additional trading partners, so as to increase her bargaining power with existing partners. But, as these incentives apply also to the other agents, it can well be that these additional investments, while costly, roughly offset each other, leading to wasteful over-expenditures in relationships that are not used in equilibrium.

---

[9] The hold-up problem is a situation where two parties may be able to work most efficiently by cooperating but refrain from doing so because of concerns that they may give the other party increased bargaining power, and thereby reduce their own profits. When an agent A has made a prior commitment to a relationship with agent B, the latter can "hold up" the former for the value of that commitment. The hold-up problem leads to severe economic cost and might also lead to underinvestment

In the case of relationship-specific investments, a hold-up problem represents a different sort of inefficiency that can lead to under-investment in relationships. Depending on the structure of the network, once the investment cost has been sunk, an agent may find not be able to recoup that sunk cost. Anticipating this possibility of being held-up, the agent may fail to invest in a relationship that would be efficient from an aggregate point of view.

A further set of inefficiencies that can arise in intermediated trade derive from the decentralized nature of decision making, which can lead to miscoordination in the routing of goods or of the prices implicit in trade terms. For example, a given buyer may value a final product only when it can be combined with the purchase of a complementary product that she buys through a different supply chain. The agents along these chains may therefore be unable to perfectly anticipate final demand for their product, as it will depend on prices and routing that take place in other paths of the network. This can lead, for example, to reduced incentives to trade, even if demand would be strong in a perfectly coordinated world.

The fact that an agent's incentive to invest in a relationship is influenced by the decisions of others is a manifestation of externalities. As noted, it is not generally clear-cut if such externalities will lead to a net under- or over-investment in relationships. But what is generally true, outside of a few highly special circumstances, is that there tends to be an inefficient set of transactions, relative to that which would maximize total value across society.

# 4    Empirical analyses of network models

When analyzing the causal determinants of behavior, where behavior is potentially influenced both by peers (connected agents) and personal characteristics, particular care must be taken. A broad observation relevant to any empirical analysis can already be seen in our exposition. In Section 2, we discussed the modeling of network formation and, in Section 3, we discussed the modeling of behaviors across a given network. In most applications, both of these dimensions are simultaneously at work: the network itself and the behaviors on that network are jointly determined by the people being studied. That is, both the network and behavior should be treated as endogenous observations, and so understanding which decisions have causal impacts on other decisions is challenging.

As friends often have similar characteristics, they are predisposed to make similar decisions, even absent any direct causal effects of one friend's decision on on another's. It is quite likely that only some of the characteristics that are relevant to the friendship decision will be observable to the analyst. For example, two students may be friends in part because they have well-educated and affluent parents (observables), but it may also be that their friendship reflects compatible personalities or attitudes towards study beyond those instilled by their parents (unobservables). As a result, if the individuals choose to become friends, and their behaviors on some dimension are correlated (e.g. both have good grades), it may not be possible able to sort out the explanation. Is it because they share the same observable characteristics (educated parents) that make them take similar actions? Is it that they face a common favorable environment (e.g., a good school and good teachers)? Or is it because the individuals had similar unobservable characteristics (attitudes towards study) that both caused them to be likely to form a friendship and, in parallel caused them to behave similarly? Disentangling such possible explanations is a major challenge that is common to empirical work dealing with the effects of social networks (e.g., see Aral and Walker, 2012). Some such studies, for instance in the enormous literature on peer effects in education, take advantage of random assignments of students to classes or as roommates (see e.g., Sacerdote, 2001; Carrell, Sacerdote and West, 2013; Algan, Dalvit, Do, Le Chapelain and Zenou, 2015). However, in many settings, we do not have the luxury of controlling the network or having it randomly assigned. Moreover, overcoming these challenges is crucial to our understanding of how networks influence (and are influenced by) behavior.

One obstacle to empirically identifying peer and social network effects is what Manski (1993) termed the *reflection problem*. If one assumes a standard linear formulation in which a person's behavior, for instance her study habits, is influenced by the average of her peers' behaviors (their study habits) and her background (her parents' education level) and her peers' backgrounds (their parents' education level), then the impact of peers' behavior on an individual's behavior cannot be distinguished from the impact of peers' background characteristics on the individual's behavior.

This problem applies most strongly when individuals are connected in groups, e.g., students into classrooms. If, instead, one has network data such that the peers who matter for some individuals differ from those who matter for some others in a sufficiently rich way, then it is possible to identify the impact of peers' behaviors, as shown by Bramoullé, Djebbari and Fortin (2009), Calvó-Armengol, Patacchini and Zenou (2009), Lee, Liu and Lin (2010), Liu and Lee (2010), Liu, Patacchini, Zenou and Lee (2012) and others.[10]

As mentioned, a different obstacle to identifying peer influences is that similar behaviors may be driven by a common factor that is not observed by the analyst. Consider, for example, a teen's decision to use illegal drugs. Is it because his friend initiated drug use, or is it due to some common influence, such as a substance-abusing parent or teacher who caused both children to adopt the same behavior? The distinction between these explanations is important for policy purposes. When peer contagion effects operate, intervening to alter one child's behavior may have the additional effect of changing several other children's behavior. In other cases, such as when children initiate substance use because adults in their neighborhood provide opportunities to do so, these so-called *multiplier effects* may not exist or may involve other factors (Brock and Durlauf, 2001). Correctly distinguishing endogenous from exogenous social effects is necessary for any effort to gauge the true net impact or social benefits of any behavioral intervention. Bramoullé, Djebbari and Fortin (2009) show how asymmetries in a peer network can be sufficient to ensure that endogenous and exogenous peer effects can be effectively identified.

---

[10] One can also identify the peer effects if the formulation is nonlinear, or with some other formulations, but then the identification is really dependent upon the deviation from linearity.

This brings us to the third major challenge outlined above: the *endogeneity of networks*.[11] To address this, one either has to have some fortuitous exogeneity in the network (Munshi, 2003; Beaman, 2012; Carrell, Sacerdote and West, 2013; Algan, Dalvit, Do, Le Chapelain and Zenou, 2015; Lindquist, Sauermann and Zenou, 2015), or one has to model the network formation directly. Recent progress has been made on this front, in a series of models of network formation that can be taken to data: e.g., Jackson and Rogers, 2007, Christakis, Fowler, Imbens and Kalyanaraman, 2010; Mele, 2013; Goldsmith-Pinkham and Imbens, 2013; Badev, 2015; Chandrasekhar and Jackson, 2014; Leung 2015; De Paula, Richards-Shubik and Tamer, 2015; Graham 2016; Patacchini, Rainone and Zenou, 2016), as well as some approaches using instrumental variables (Bifulco, Fletcher and Ross, 2011; Patacchini and Zenou, 2016).[12]

# 5    Concluding remarks

There are many important research frontiers in social network analysis, especially at junctures between sociology, economics, anthropology and other fields. These frontiers include building a richer understanding of the dynamics of networked relationships and behaviors, and their co-evolution.

Perhaps most importantly, as social network analysis is increasingly making its way into policy arenas, it is important to understand potential inefficiencies both in network structure and in the resulting behaviors that they drive. Our main emphasis in this survey has been that the concept of externalities is a unifying principle that can be incorporated into network analysis more generally, as it helps frame many networked interactions and is the defining characteristic of an interesting network. In the absence of externalities one could simply examine bilateral relationships independently of each other and divorced from their social context. What makes the social context interesting is that a given relationship is influenced by, and influences, the more general social structure. Viewing that context through the lens of the underlying externalities focuses our attention on exactly how peoples' behaviors are interrelated and what is it that drives inefficiencies and imperfections in human behaviors and societies.

---

[11] Unfortunately, the sort of instrumental variable strategy described above is valid only if the network can be treated as (conditionally) exogenous, which is not usually the case unless one has access to an appropriately controlled field experiment, not only in terms of observed characteristics but also in terms of unobserved characteristics that could influence behaviors. It could then be that behaviors of friends of friends are correlated in ways that invalidate instrumenting on friends of friends as a method to identify peer effects.

[12] We refer to the literature surveys provided by Blume, Brock, Durlauf and Ioannides, 2011, Jackson (2014), Graham (2015), Jackson, Rogers and Zenou (2016), Chandrasekhar (2016), Fortin and Boucher (2016) for a detailed treatment of these econometric issues.

Beyond the importance of externalities, there are increasingly rich network analyses involving economic applications ranging from development economics (e.g., Karlan et al 2009,2010, Munshi and Rosenzweig 2009, Kinnan and Townsend 2012, Banerjee et al 2013, Feigenberg, Field, and Pande 2013, Cai, de Janvry, and Sadoulet 2015, Breza 2016 ), to labor economics (e.g., Montgomery 1992, Calvo-Armengal and Jackson 2004, Ioannides and Datcher-Loury 2007, Beaman 2016), to international trade (e.g., Chaney 2016), to interstate alliances and wars (e.g., Jackson and Nei, 2015, Koenig et al 2016), politics (e.g., Cohen and Malloy, 2014), and crime (e.g., Glaeser, Sacerdote, and Scheinkman 1996, Calvó-Armengol and Zenou, 2004; Zenou, 2016). Theoretical work is proceeding in tandem with empirical studies across these, and other, varied contexts, which makes clear that the interests of economists, sociologists and other social scientists will continue to overlap in substantial ways. As this research progresses, and as our data becomes ever more complete and integrated, it will become increasingly important to marry the insights from different scientific fields to gain a more complete understanding of social context and how it drives human behavior.